# Debye temperature in LaH$_x$-LaD$_y$ superconductors

E. F. Talantsev[1,2*]

[1]M.N. Mikheev Institute of Metal Physics, Ural Branch, Russian Academy of Sciences, 18, S. Kovalevskoy St., Ekaterinburg, 620108, Russia

[2]NANOTECH Centre, Ural Federal University, 19 Mira St., Ekaterinburg, 620002, Russia

[*]E-mail: evgeny.talantsev@imp.uran.ru

*Abstract*

The Debye temperature, $T_\theta = \frac{\hbar}{k_B} \cdot \omega_\theta$, is a derivative of the Debye frequency, $\omega_\theta$, is integrated characteristic frequency of full phonon spectrum, $\alpha^2(\omega)F(\omega)$. In the BCS theory, $T_\theta$ in conjunction with electron-phonon coupling strength parameter, $\lambda_{e-ph}$, determines the superconducting transition temperature, $T_c$. Despite a fact that more accurate theory of electron-phonon mediated superconductivity requires the knowledge of full phonon spectrum, $\alpha^2(\omega)F(\omega)$, which can be very accurately computed by first principles calculation technique, there is no experimental technique which can measure $\alpha^2(\omega)F(\omega)$ in highly-compressed near-room-temperature (NRT) superconductors. Thus, $\omega_\theta$ remains to be the only measurable parameter of full phonon spectrum, $\alpha^2(\omega)F(\omega)$, which can be deduced by the fit of experimental temperature-dependent resistance data, $R(T)$, to Bloch-Grüneisen equation. Taking in account that within electron-phonon mediated theory of superconductivity two isotopic counterparts (or, in case of NRT superconductors, the same superconducting phase at different pressures), designated by subscripts of 1 and 2, should be obey the relation of $T_{\theta 1}/T_{\theta 2}=T_{c1}/T_{c2}$, there is a way to reaffirm/disprove the electron-phonon mechanism of NRT superconductivity. In this paper, we perform the analysis for *R3m*-phase of H$_3$S at different pressure, as well as for several superconductors in LaH$_x$-LaD$_y$ system and show that there is a large disagreement between experimental data and $T_{\theta 1}/T_{\theta 2}=T_{c1}/T_{c2}$. Taking in account that similar disagreement has recently reported in $H_3S$-$D_3S$ system, it can be concluded that primary origin for NRT superconductivity remains to be discovered.

**Debye temperature in $LaH_x$-$LaD_y$ superconductors**

**1. Introduction**

Highly-compressed lanthanum decahydride exhibits the highest reported to date superconducting transition temperature of $T_c \gtrsim 240\ K$ [1,2]. There is a widely accepted point of view [3] that near-room-temperature (NRT) superconductivity in $LaH_{10}$ and other highly-compressed hydrogen-rich compounds is originated from the electron-phonon coupling mechanism proposed by Bardeen, Cooper and Schrieffer (BCS) [4]. The strength of the electron-phonon coupling is quantified in the Eliashberg theory [5] by the coupling strength constant, $\lambda_{e-ph}$:

$$\lambda_{e-ph} = 2 \cdot \int_0^\infty \frac{\alpha^2(\omega) \cdot F(\omega)}{\omega} \cdot d\omega \tag{1}$$

where ω is the phonon frequency, $F(\omega)$ is the phonon density of states, and $\alpha^2(\omega) \cdot F(\omega)$ is the electron-phonon spectral function (more details can be found elsewhere [6-10]).

It should be stressed, however, that the majority of applications of the BCS and the Eliashberg theories to predict NRT superconductivity in hydrogen-rich highly compressed compounds have been failed. For instance, we can mention the case of highly-compressed silane, $SiH_4$, for which $T_c \cong 98 - 107\ K$ was predicted by Li *et al* [11] and $T_c = 166\ K$ was predicted by Aschroft's group [12], while the experiment showed $T_c \leq 13\ K$ so far [13].

Based on this, more thoroughly theoretical analysis of recent experimental milestone discoveries of NRT superconductivity in some hydride compounds [14-19] is required, because each successfully and each unsuccessfully predicted NRT superconductor cases should be treated with equal weight within total database, without fixing the database size by pre-defined stopping rule for unsuccessful cases. This approach will advance the progress in the field because it eliminates so-called survivorship bias [20,21].

One of the way to advance theoretical understanding of NRT superconductivity is to continue to test the validity of electron-phonon coupling mechanism as potential origin for NRT superconductivity in highly-compressed hydrides/deuterides. In this paper, we deduce the Debye temperature, $T_\theta$, for superconductors in $LaH_{10}$-$LaD_{10}$ system with the purpose to reaffirm/disprove the following theoretical derivative of the electron-phonon pairing mechanism [22]:

$$\left.\frac{T_{c,1}}{T_{c,2}}\right|_{exp} = \left.\frac{\omega_{ln,1}}{\omega_{ln,2}}\right|_{first-principles\ calcs} = \left.\frac{T_{\theta,1}}{T_{\theta,2}}\right|_{exp}, \quad (2)$$

where subscripts of 1 and 2 indicate two isotopic counterparts, and $\omega_{ln}$ is logarithmic phonon frequency given by:

$$\omega_{ln} = exp\left[\frac{\int_0^\infty \frac{ln(\omega)}{\omega} \cdot F(\omega) \cdot d\omega}{\int_0^\infty \frac{1}{\omega} \cdot F(\omega) \cdot d\omega}\right] \quad (3)$$

where $\omega$ is the phonon frequency, $F(\omega)$ is the phonon density of states, and $\alpha^2(\omega) \cdot F(\omega)$ is the electron-phonon spectral function (more details can be found elsewhere [6-10]).

In result, we show that superconductors in $LaH_{10}$-$LaD_{10}$ system do not comply with Eq. 1, and this alludes that alternative pairing mechanisms [23-27] which causes the rise of NRT superconductivity in $LaH_{10}$-$LaD_{10}$ system need to be considered.

## 2. Problem associated with Allen-Dynes model

Within electron-phonon mediated phenomenology of superconductivity all materials can be characterized as weak ($\lambda_{e-ph} \ll 1$), intermediate ($\lambda_{e-ph} \sim 1$), and strong ($\lambda_{e-ph} \gg 1$) coupled superconductors. For weak-coupled superconductors Bardeen, Cooper and Schrieffer [4] derived an expression which links $T_c$, $T_\theta$ and $\lambda_{e-ph}$:

$$T_c = T_\theta \cdot e^{-\left(\frac{1}{\lambda_{e-ph} - \mu^*}\right)} \quad (4)$$

where μ* is the Coulomb pseudopotential parameter, which is within a range of μ* = 0.10-0.17 [6-10].

McMillan [6] performed advanced analysis of the problem and proposed an equation:

$$T_c = \left(\frac{1}{1.45}\right) \cdot T_\theta \cdot e^{-\left(\frac{1.04 \cdot (1+\lambda_{e-ph})}{\lambda_{e-ph}-\mu^* \cdot (1+0.62 \cdot \lambda_{e-ph})}\right)} \tag{5}$$

which covers a wide range of the coupling strength of $\mu^* \leq \lambda_{e-ph} \leq 1.65$ [8].

And one of the most widely used equation in the field was proposed by Allen and Dynes [8]:

$$T_c = \left(\frac{1}{1.20}\right) \cdot \left(\frac{\hbar}{k_B}\right) \cdot \omega_{ln} \cdot e^{-\left(\frac{1.04 \cdot (1+\lambda_{e-ph})}{\lambda_{e-ph}-\mu^* \cdot (1+0.62 \cdot \lambda_{e-ph})}\right)} \cdot f_1 \cdot f_2 \tag{6}$$

where:

$$f_1 = \left(1 + \left(\frac{\lambda_{e-ph}}{2.46 \cdot (1+3.8 \cdot \mu^*)}\right)^{3/2}\right)^{1/3} \tag{7}$$

$$f_2 = 1 + \frac{\left(\frac{\langle\omega^2\rangle^{1/2}}{\omega_{ln}} - 1\right) \cdot \lambda_{e-ph}^2}{\lambda_{e-ph}^2 + \left(1.82 \cdot (1+6.3 \cdot \mu^*) \cdot \left(\frac{\langle\omega^2\rangle^{1/2}}{\omega_{ln}}\right)\right)^2}, \tag{8}$$

$$\langle\omega^2\rangle^{1/2} = \frac{2}{\lambda_{e-ph}} \cdot \int_0^\infty \omega \cdot \alpha^2(\omega) \cdot F(\omega) \cdot d\omega. \tag{9}$$

where $f_1$ and $f_2$ are so-called the strong-coupling correction function and the shape correction function, respectively [8]. If $f_2$ function (Eq. 8) can be approximated by simple parabolic analytical expression [22]:

$$f_2^* = 1 + (0.0241 - 0.0735 \cdot \mu^*) \cdot \lambda_{e-ph}^2. \tag{10}$$

(the necessity of this approximation is due to originally defined $f_2$ function (Eq. 8) requires the measurement of full phonon spectrum, $\alpha^2(\omega) \cdot F(\omega)$, which is challenging experimental task for highly-compressed superconductors), the logarithmic phonon frequency, $\omega_{ln}$ (Eq. 3), defined by Allen and Dynes [7,8] has severe fundamental problem on its definition. Truly, if one can consider the integrand part in the square brackets in the left part of the $\omega_{ln}$ definition:

$$\omega_{ln} = exp\left[\frac{\int_0^\infty \left[\frac{ln(\omega)}{\omega}\right] \cdot F(\omega) \cdot d\omega}{\int_0^\infty \frac{1}{\omega} \cdot F(\omega) \cdot d\omega}\right] = exp\left[\frac{\int_0^\infty \left[ln\left((\omega)^{\frac{1}{\omega}}\right)\right] \cdot F(\omega) \cdot d\omega}{\int_0^\infty \frac{1}{\omega} \cdot F(\omega) \cdot d\omega}\right]. \quad (11)$$

then the logarithm part of it:

$$ln(\omega) \quad (12)$$

cannot be accepted to have physical meaning because as any other functions (i.e., $exp(x)$, $cos(x)$, modified Bessel function of $K_n(x)$, etc.), the logarithm can be taken only from unit-less variable, but ω has the unit of Hz. All physical laws where oscillations are primary variable (for instance, Rayleigh's scattering law, Planck's law, Fourier transformation, etc.) utilize this physical value, but it has multiplicative numerators which eliminate the Hz unit.

However, due to all first principles calculations papers (published to date) for highly-compressed hydrogen-rich superconducting system utilize Eqs. 6-9 to compute $T_c$, we will use the ratio of computed pair of $\omega_{ln,LaH10}$ and $\omega_{ln,LaD10}$ to test the validity of Eq. 2.

Based on all above, advanced McMillan equation [22]:

$$T_c = \left(\frac{1}{1.45}\right) \cdot T_\theta \cdot e^{-\left(\frac{1.04 \cdot \left(1+\lambda_{e-ph,aMcM}\right)}{\lambda_{e-ph}-\mu^* \cdot \left(1+0.62 \cdot \lambda_{e-ph,aMcM}\right)}\right)} \cdot f_1 \cdot f_2^* \quad (13)$$

where $\lambda_{e-ph,aMcM}$ is the electron-phonon coupling strength constant, represents physically backgrounded law, which exhibits excellent accuracy for $\mu^* < \lambda_{e-ph,aMcM} \lesssim 1.65$.

To deduce the Debye temperature, $T_\theta$, for highly-compressed LaH$_{10}$ and LaD$_{10}$ superconductors we employ the fit of experimental temperature dependent resistance data, $R(T)$ (or reduced resistance data), to Bloch-Grüneisen (BG) equation [28,29]:

$$R(T) = R_0 + A \cdot \left(\frac{T}{T_\theta}\right)^5 \cdot \int_0^{\frac{T_\theta}{T}} \frac{x^5}{(e^x-1) \cdot (1-e^{-x})} \cdot dx \quad (14)$$

where, the first term represents residual resistance arises from the scattering of conduction charge carriers on the static defects of crystalline lattice, while the second term describes the charge scattering due to the interaction with phonons, and $A$ and $T_\theta$ are free-fitting parameters.

Raw experimental $R(T)$ data for LaH$_{10}$ and LaD$_{10}$ superconductors was kindly provided by Dr. M. I. Eremets and Dr. V. S. Minkov (Max-Planck Institut für Chemie, Mainz, Germany) and data for *R3m* phase of highly-compressed H$_3$S superconductor by M. Einaga (Osaka University, Japan).

## 3. Results and Discussion

Guigue *et al* [30] synthesized pure H$_3$S phase by laser heating hydrogen-embedded solid sulphur samples at pressures above 75 GPa. Diffraction studies showed that the compound has the crystal structure with space group of *Cccm* which exhibits up to pressure of $P = 160$ GPa. It should be noted, that *Cccm* phase of H$_3$S is non-superconducting. In contrast, Einaga *et al* [31] reported that H$_3$S compound synthesized from gaseous H$_2$S has low-pressure ($P \leq 150\ GPa$) low-$T_c$ phase with space group of *R3m*, and high-pressure ($P > 150\ GPa$) high-$T_c$ phase with space group of *Im-3m*.

Most extensive study for the phase transitions in highly-compressed sulphur hydride when gaseous H$_2$S is used as a precursor was reported by Goncharov *et al.* [32] who found a rich homological series of sulphur hydride phases, H$_n$S$_m$, which form at high-pressure and laser annealing conditions. Thus, phase composition/phase symmetry studies for highly-compressed sulphur hydride are at ongoing stage and the agreement between research groups be reached in a future, here in Section 3.1 we report results on the evolution of $T_\theta$ and $\lambda_{e-ph,aMcM}$ vs applied pressure of $111\ GPa \leq P \leq 150\ GPa$ for low-$T_c$ *R3m*-superconducting phase of H$_3$S compound reported by Einaga *et al* [31].

Drozdov *et al* [16] reported large sets of experimental $R(T)$ curves for superconductors in LaH$_{10}$ and LaD$_{10}$ system exhibiting different hydrogen/deuterium stoichiometry.

Unfortunately, $R(T)$ curves for Sample 1 (LaH$_{10}$, $T_c \cong 250\ K$) and Sample 17 (LaD$_{10}$, $T_c \cong 150\ K$) showed in Fig. 4 [16] have very short normal state part of $R(T)$ curve, that deduced Debye temperatures, $T_\theta$, for these two samples have large uncertainties which exceed the deduced $T_\theta$. However, due to Drozdov *et al* [16] reported so extensive studies of LaH$_{10}$-LaD$_{10}$ system, that some exemptions do not significantly limited reported results herein.

### 3.1. *R3m* phase of H$_3$S

Einaga *et al.* [31] in their Fig. 3(a) reported $R(T)$ curves for H$_3$S measured at applied pressure in the range of $111\ GPa \leq P \leq 150\ GPa$. All reported $R(T)$ curves reach $R = 0\ \Omega$ point, however samples subjected to pressures of $P = 133$ GPa and 150 GPa have long tails to reach zero resistance with inflection points at $R(T)/R_{norm}(T) \sim 0.05$, where $R_{norm}(T)$ is extrapolated curve of $R(T)$ fit to BG equation (Eq. 15). Based on this, $T_c$ is defined by $R(T)/R_{norm}(T) = 0.05$ criterion for H$_3$S samples considered in this Section.

At pressure range of $111\ GPa \leq P \leq 150\ GPa$ annealed H$_3$S compound exhibits in *R3m* phase. Taking in account that experiment [31] shows that $T_c$ is monotonically changing vs applied pressure there is an expectation that $T_\theta$ and $\lambda_{e\text{-}ph}$ will be also following monotonic trends. However, the analysis of experimental $R(T)$ data (Fig. 1, Table 1) shows that the $T_\theta$ and $\lambda_{e\text{-}ph}$ are varying in a random way in comparison with $T_c$ (Fig. 2), which is an evidence that in *R3m* phase of H$_3$S, one of integrated characteristic of the phonon spectrum, which is $T_\theta$, does not correlate with the superconducting transition temperature, $T_c$.

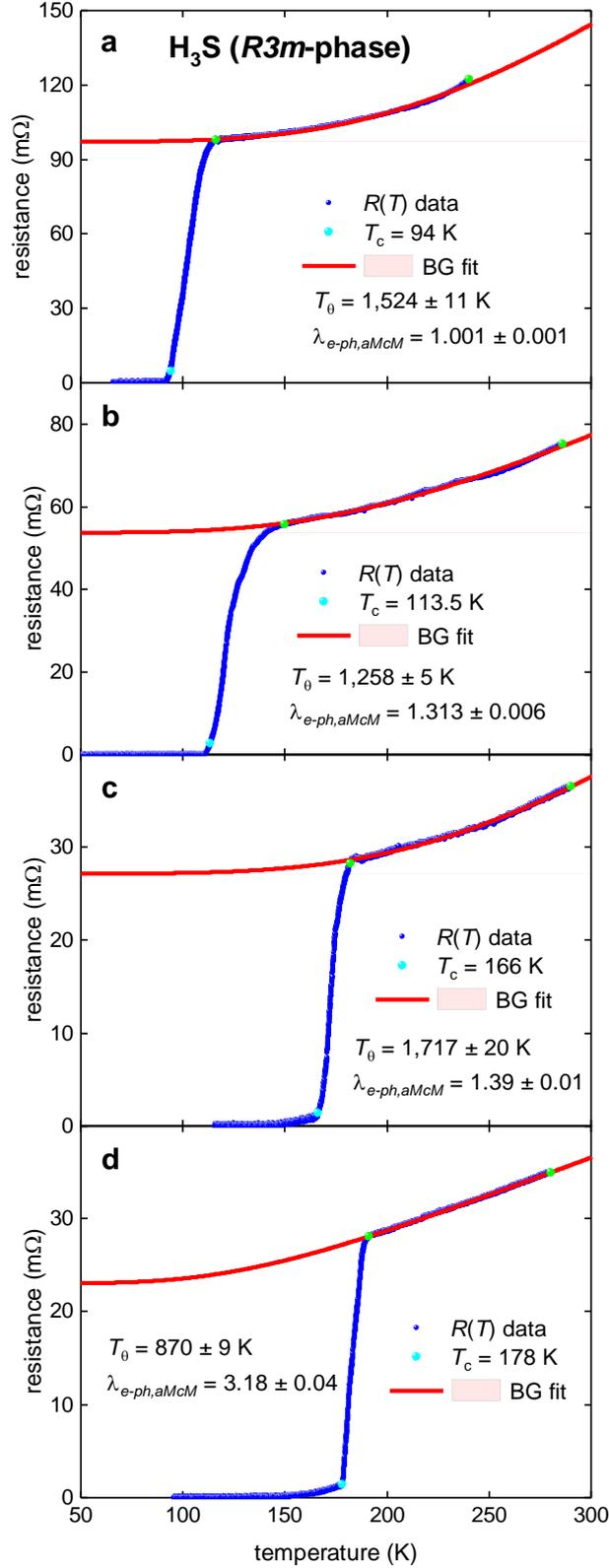

**Figure 1.** *R*(*T*) data and fit to BG model for highly-compressed *R3m*-phase of H$_3$S ($111\ GPa \leq P \leq 150\ GPa$). Raw *R*(*T*) data is from Ref. [31]. 95% confidence bars are shown. Green balls indicate the bounds for which *R*(*T*) data was used for the fit. Cyan ball shows $T_c$ defined by the $R(T)/R_{norm}(T) = 0.05$ criterion. Showed $\lambda_{e-ph,aMcM}$ values calculated for µ* = 0.10. (a) Fit quality $R = 0.997$; (b) $R = 0.998$; (c) $R = 0.998$; (d) $R = 0.9993$.

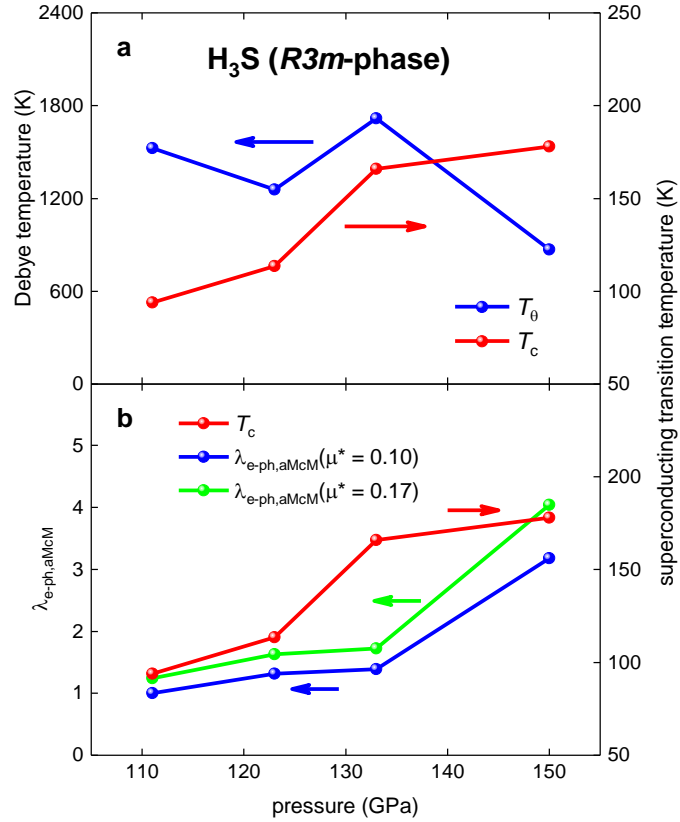

**Figure 2.** (a) Deduced Debye temperatures, $T_\theta$, and superconducting transition temperature, $T_c$, for highly-compressed *R3m*-phase of $H_3S$ ($111\ GPa \leq P \leq 150\ GPa$) vs applied pressure. (b) Superconducting transition temperature, $T_c$, and calculated electron-phonon coupling strength constant for highly-compressed *R3m*-phase of $H_3S$ ($111\ GPa \leq P \leq 150\ GPa$) vs applied pressure. Raw data is from Ref. 31.

**Table 1.** Deduced $T_\theta$ and calculated $\lambda_{e\text{-ph}}$ for highly-compressed *R3m*-phase of $H_3S$ at $111\ GPa \leq P \leq 150\ GPa$. $T_c$ values are defined by $R(T)/R_{norm}(T) = 0.05$ criterion.

| Pressure (GPa) | $T_\theta$ (K) | $T_c$ (K) | Assumed $\mu^*$ | $\lambda_{e-ph,BCS}$ | $\lambda_{e-ph,aMcM}$ | $\lambda_{e-ph}$ (first-principles calculations) [33] |
|---|---|---|---|---|---|---|
| 111 | $1524 \pm 11$ | 94.0 | 0.10 | $0.459 \pm 0.001$ | $1.001 \pm 0.004$ | |
| | | | 0.17 | $0.529 \pm 0.001$ | $1.241 \pm 0.006$ | |
| 123 | $1258 \pm 5$ | 113.5 | 0.10 | $0.516 \pm 0.001$ | $1.313 \pm 0.006$ | |
| | | | 0.17 | $0.586 \pm 0.001$ | $1.628 \pm 0.006$ | 2.07 |
| 133 | $1717 \pm 20$ | 166.0 | 0.10 | $0.528 \pm 0.002$ | $1.391 \pm 0.013$ | |
| | | | 0.17 | $0.598 \pm 0.002$ | $1.726 \pm 0.017$ | |
| 150 | $870 \pm 9$ | 178.0 | 0.10 | $0.730 \pm 0.002$ | $3.18 \pm 0.04$ | |
| | | | 0.17 | $0.805 \pm 0.003$ | $4.04 \pm 0.06$ | |

It should be noted that deduced $\lambda_{e\text{-ph,aMcM}}$ values (Table 1) cover so wide range of $1.00 \leq \lambda_{e-ph,aMcM} \leq 4.03$ that there is no possibility to affirm/disprove theoretical value of $\lambda_{e-ph,aMcM} = 2.07$ reported for *R3m*-phase by Duan *et al* [33]. However, there is another way to utilize Eqs. 2,15, because ones do not only apply for two isotopic counterparts, but also can be apply for the same compound at the same phase state when the superconducting transition temperature of the material is changing vs the change in the pressure:

$$\left.\frac{T_{c,n}}{T_{c,m}}\right|_{exp} = \left.\frac{T_{\theta,n}}{T_{\theta,m}}\right|_{exp}, \quad (15)$$

where the subscript *m* indicates a *m*-stage of the compression and *n* indicates *n*-stage of the compression. If the NRT superconductivity is mediated by the electron-phonon interaction, then Eq. 15 should be valid. In Fig. 3 we show data for *R3m*-phase of $H_3S$ for ratios of:

$$\left.\frac{T_{c,n}}{T_{c,P=111\,GPa}}\right|_{exp} = \left.\frac{T_{\theta,n}}{T_{\theta,P=111\,GPa}}\right|_{exp}, \quad (16)$$

Results (Fig. 3) are in a large disagreement with the assumption that the NRT superconductivity is mediated by the electron-phonon mechanism in *R3m*-phase of $H_3S$. The most compelling case is for pressure of *P* = 150 GPa, where the disagreement between expected and deduced (from experiment) values is in more than three times.

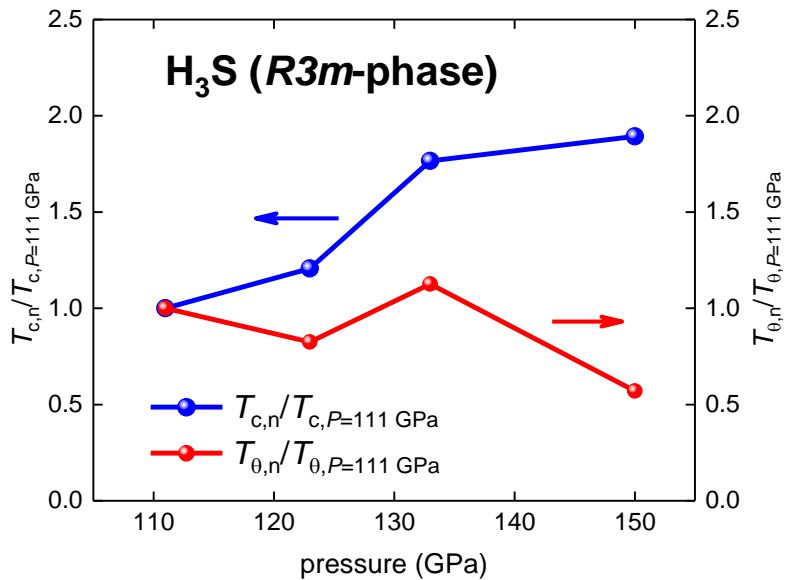

**Figure 3.** The ratios of $\frac{T_{c,n}}{T_{c,P=111\,GPa}}$ and $\frac{T_{\theta,n}}{T_{\theta,P=111\,GPa}}$ for highly-compressed *R3m*-phase of $H_3S$ ($111\,GPa \leq P \leq 150\,GPa$) vs applied pressure.

### 3.2. Superconductors in $LaH_{10}$-$LaD_{10}$ system

#### 3.2.1. $LaH_{10}$ with $T_c$ = 240 K

We start our analysis of $LaH_x$-$LaD_y$ system by the analysis of $R(T)$ dataset for Sample 3 which exhibits the highest transition temperature of $T_c$ = 240 K in the report by Drozdov *et al* [16] in their Fig. 2. This sample has two inflection points in $R(T)$ curves which can be seen in in Fig. 4, and we calculate $\lambda_{e-ph,aMcM}$ values for these two superconducting transition temperatures (Table 2). In our calculations we use µ* = 0.10 reported by Errea *et al.* [34].

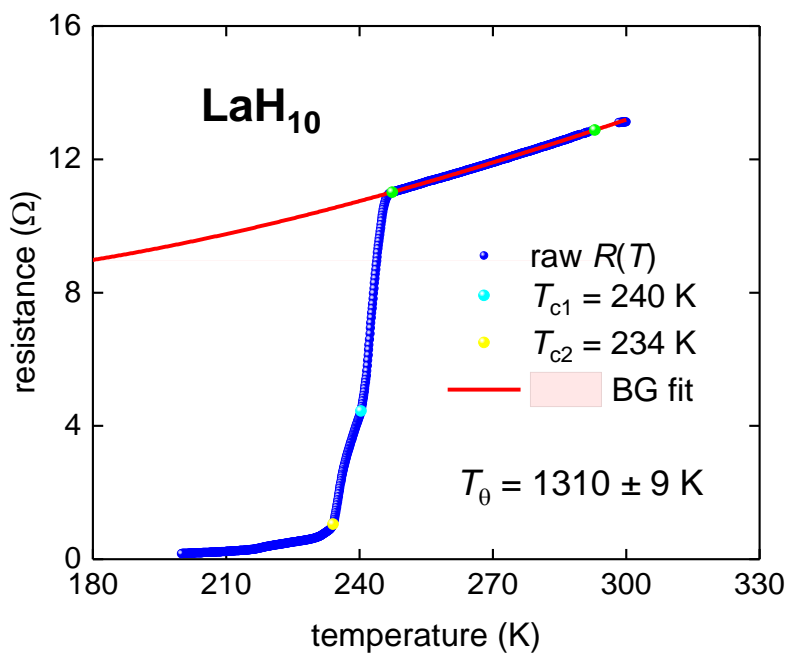

**Figure 4.** $R(T)$ data and fit to BG model for highly-compressed $LaH_{10}$ (Sample 3 [16]). Raw $R(T)$ data is from Ref. [16]. 95% confidence bars are shown. Green balls indicate the bounds for which $R(T)$ data was used for the fit. Cyan and yellow balls show $T_c$ defined by two inflection points. Fit quality $R$ = 0.99992.

**Table 2.** Deduced $T_\theta$ and calculated $\lambda_{e-ph,aMcM}$ for highly-compressed $LaH_{10}$ (Sample 3 [16]). Assumed µ* = 0.10 [34]. $T_c$ values defined by the inflection points in $R(T)$ curve.

| Compound (Sample ID; pressure) | $T_\theta$ (K) | $T_c$ (K) | $\lambda_{e-ph,BCS}$ | $\lambda_{e-ph,aMcM}$ | $\lambda_{e-ph}$ (first-principles calculations) [34] |
|---|---|---|---|---|---|
| $LaH_{10}$ (Sample 3; $P$ = 150 GPa) | 1310 ± 9 | 240 | 0.689 ± 0.002 | 2.77 ± 0.02 | 2.76 ($P$ = 163 GPa) |
| | | 234 | 0.681 ± 0.002 | 2.69 ± 0.02 | |

One can see an excellent agreement between computed (by first principles calculations [34]) and deduced (by our analysis herein) the electron-phonon coupling strength values, $\lambda_{e-ph}$, for LaH$_{10}$ compound (Table 2). However, the rest of available experimental datasets shows the large disagreement between computed and deduced (from experiment) values, which we present below. And these majority of disagreement case, as we mentioned above, should be considered with equal weight with successful cases, to understand the nature of NRT superconductivity in highly-compressed super-hydrides/deuterides.

### 3.2.2. LaH$_x$ with $T_c$ ~ 215 K

Drozdov *et al* [16] in their Extended Data Fig. 5 reported $R(T)/R_{norm}(T)$ curve for laser annealed LaH$_x$ (Sample 12) with very sharp superconducting transition with $T_c$ ~ 210 K at $P$ = 160 GPa. When the pressure was decreased to $P$ = 150 GPa, the transition temperature increased to $T_c$ ~ 215 K. Reduced resistance curve, $R(T)/R_{norm}(T)$, at $P$ = 150 GPa is analysed in Fig. 5 with deduced parameters show in Table 3.

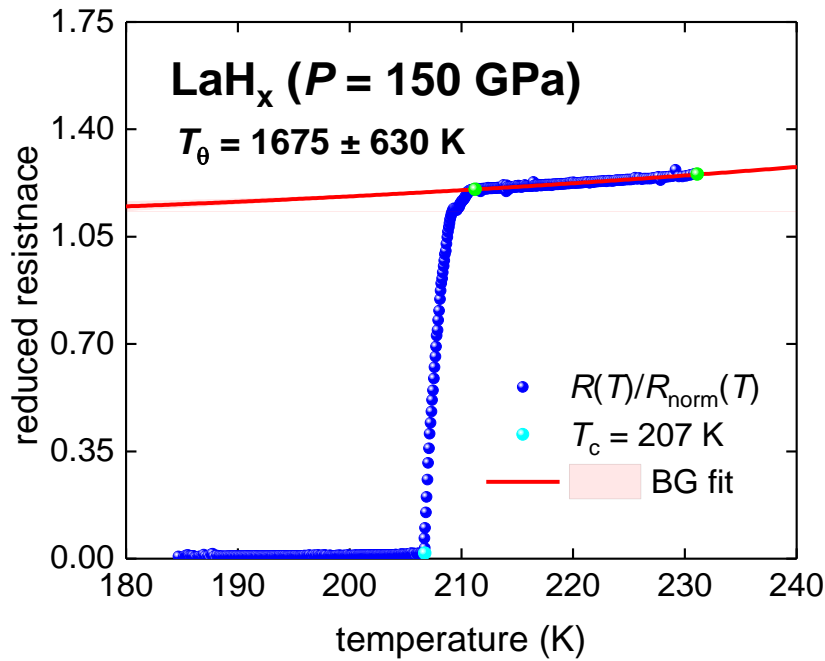

**Figure 5.** $R(T)/R_{norm}(T)$ data and fit to BG model for highly-compressed LaH$_{10}$ (Sample 12 [16]). 95% confidence bars are shown. Green balls indicate the bounds for which $R(T)$ data was used for the fit. Cyan ball shows $T_c$ defined by the $R(T)/R_{norm}(T) = 0.01$ criterion. Fit quality $R = 0.959$.

**Table 3.** Deduced $T_\theta$ and calculated $\lambda_{\text{e-ph,aMcM}}$ for highly-compressed LaH$_x$ (Sample 12 [16]). Assumed $\mu^* = 0.10$ [34]. $T_c$ value defined by zero resistance point.

| Compound (Sample ID; pressure) | $T_\theta$ (K) | $T_c$ (K) | $\lambda_{e-ph,BCS}$ | $\lambda_{e-ph,aMcM}$ | $\lambda_{e-ph}$ (first-principles calculations) [34] |
|---|---|---|---|---|---|
| LaH$_x$ (Sample 12; $P$ = 150 GPa) | 1675 ± 630 | 207 | 0.58 ± 0.14 | $1.75^{+1.4}_{-0.4}$ | 2.67-3.62 |

Due to the normal part of the resistance curve for this sample is relatively narrow, the uncertainty in deduced Debye temperature is large, however, calculated value of $\lambda_{e-ph,aMcM} = 1.75^{+1.4}_{-0.4}$ seems to be still far apart from computed value range of $\lambda_{e-ph,aMcM} = 2.67 - 3.62$ reported by Errea *et al* [34].

### 3.2.3. LaH$_x$ and LaD$_y$ with $T_c \sim$ 65 K

Drozdov *et al* [16] in their Extended Data Fig. 5 reported $R(T)/R_{\text{norm}}(T)$ curves for LaH$_x$ (Sample 11) and LaD$_y$ (Sample 14) with very close superconducting transition temperatures. By use of the $R(T)/R_{\text{norm}}(T) = 0.05$ criterion, the transition temperatures are found to be $T_c$ = 66.2 K and $T_c$ = 65.3 K for LaH$_x$ and LaD$_y$ respectively. This is practically ideal pair to test the validity of electron-phonon mediated NRT superconductivity in LaH-LaD system, because the ratio of transition temperatures for these isotopic counterparts is practically undistinguishable from the unity:

$$\left.\frac{T_{c,2}}{T_{c,1}}\right|_{exp} = \frac{66.2\ K}{65.3\ K} = 1.014 \cong 1.0 \qquad (16)$$

where subscripts 1 and 2 designate LaH$_x$ and LaD$_y$ compounds respectively.

In Fig. 5 and Table 3 we show temperature dependent of the reduced resistance and data fits to BG model. Deduced ratio for the Debye temperatures for these isotopic counterparts is:

$$\left.\frac{T_{\theta,2}}{T_{\theta,1}}\right|_{exp} = \frac{603\ K}{415\ K} = 1.453 \cong 1.5 \qquad (17)$$

which is remarkably different from the ratio of the transition temperatures (Eq. 17). From this we can conclude there is no experimental evidences that electron-phonon mechanism is the origin for superconductivity in these low-$T_c$ samples of $LaH_x$ and $LaD_y$.

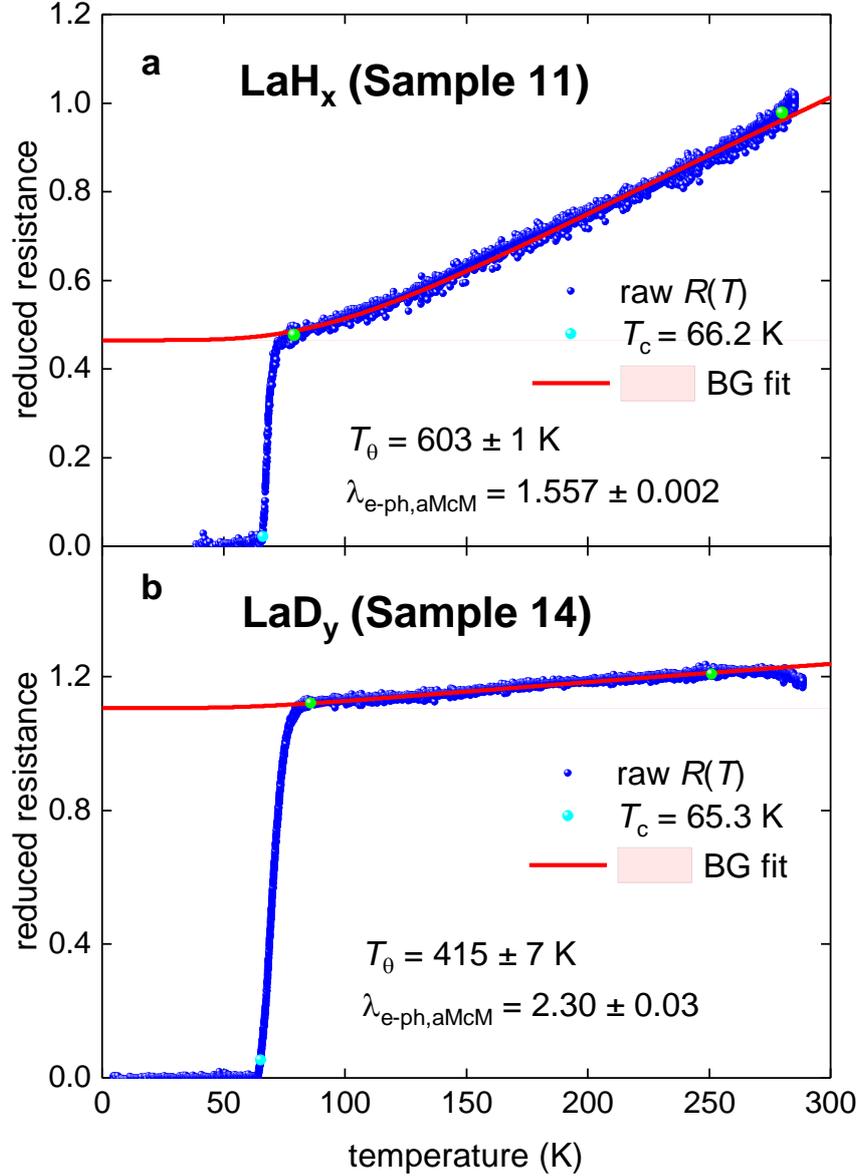

**Figure 6.** $R(T)$ data and fit to BG model for highly-compressed $LaH_{10}$ (Sample 11 [16], panel **a**) and $LaD_{10}$ (Sample 14 [16], panel **b**). Raw $R(T)$ data is from Ref. [16]. 95% confidence bars are shown. Green balls indicate the bounds for which $R(T)$ data was used for the fit. Cyan ball shows $T_c$ defined by the $R(T)/R_{norm}(T) = 0.05$ criterion. (a) Fit quality $R = 0.997$; (b) $R = 0.982$.

**Table 4.** Deduced $T_\theta$ and calculated $\lambda_{\text{e-ph,aMcM}}$ for highly-compressed LaH$_x$ (Sample 11 [Dr]) and LaD$_y$ (Sample 14 [Dr]). Assumed $\mu^* = 0.10$ [34]. $T_c$ values defined by the inflection points in $R(T)$ curve. $T_c$ values are defined by $R(T)/R_{\text{norm}}(T) = 0.05$ criterion.

| Compound (Sample ID; pressure) | $T_\theta$ (K) | $T_c$ (K) | $\lambda_{e-ph,BCS}$ | $\lambda_{e-ph,aMcM}$ | $\lambda_{e-ph}$ (first-principles calculations) [34] |
|---|---|---|---|---|---|
| LaH$_x$ (Sample 11; $P = 150$ GPa) | 603 ± 1 | 66.2 | 0.552 ± 0.001 | 1.557 ± 0.002 | 2.67-3.62 |
| LaD$_y$ (Sample 14; $P = 130$ GPa) | 415 ± 7 | 65.3 | 0.641 ± 0.002 | 2.30 ± 0.03 | 3.14 |

### 3.2.4. LaH$_x$ and LaD$_{11}$ with $T_c \sim 100$ K

Drozdov *et al* [16] in their Extended Data Fig. 5 reported $R(T)/R_{\text{norm}}(T)$ curves for different laser annealing stage of LaH$_x$ (Sample 10) and LaD$_{11}$ (Sample 8) specimens which have superconducting transition temperatures near 100 K, if the transition will be defined by the inflection point (for Samples in Fig. 7,a and 7,c) or by the for criterion of $R(T)/R_{\text{norm}}(T) = 0.25$ (Fig. 7,b). Taking in account that $R(T)/R_{\text{norm}}(T)$ curve for LaD$_{11}$ (Sample 8, Fig. 7,b) has broad low-temperature tail with clearly observed inflection point at $T = 125$ K for which the $T_c$ criterion is $R(T)/R_{\text{norm}}(T) = 0.25$, the same $T_c$ criterion was applied for laser annealed LaH$_x$ counterpart (Sample 10, Fig. 7,b), for which the transition temperature defines as $T_c = 107$ K (Fig. 7,b).

Thus, in Table 6 we calculated $\lambda_{\text{e-ph,aMcM}}$ values for these three compounds with deduced $T_\theta$ from the fit of $R(T)/R_{\text{norm}}(T)$ curve to BG equation which are shown in Fig. 7.

Laser-annealed isotopic counterparts LaH$_x$ (Sample 10, Fig. 6,b) and LaD$_{11}$ (Sample 8) have reasonably close ratios:

$$\left.\frac{T_{c,2}}{T_{c,1}}\right|_{exp} = \frac{125\ K}{107\ K} = 1.17 \neq \left.\frac{T_{\theta,1}}{T_{\theta,2}}\right|_{exp} = \frac{1199\ K}{941\ K} = 1.27 \qquad (18)$$

where subscripts 1 and 2 designate LaH$_x$ (Sample 10, Fig. 7,b) and LaD$_{11}$ compounds respectively. It should be noted, that for this NRT pair, $T_c$ and $T_\theta$ for hydrogen-based compound (i.e. LaH$_x$) are lower than ones for deuterium-based compound. In overall, computed $\lambda_{\text{e-ph}}$ values by Errea *et al.* [34] in assumption of $\mu^* = 0.10$ (Table 5) for these NRT superconductors are very different from deduced $\lambda_{e-ph}$ value we deduced in our analysis herein (Table 5).

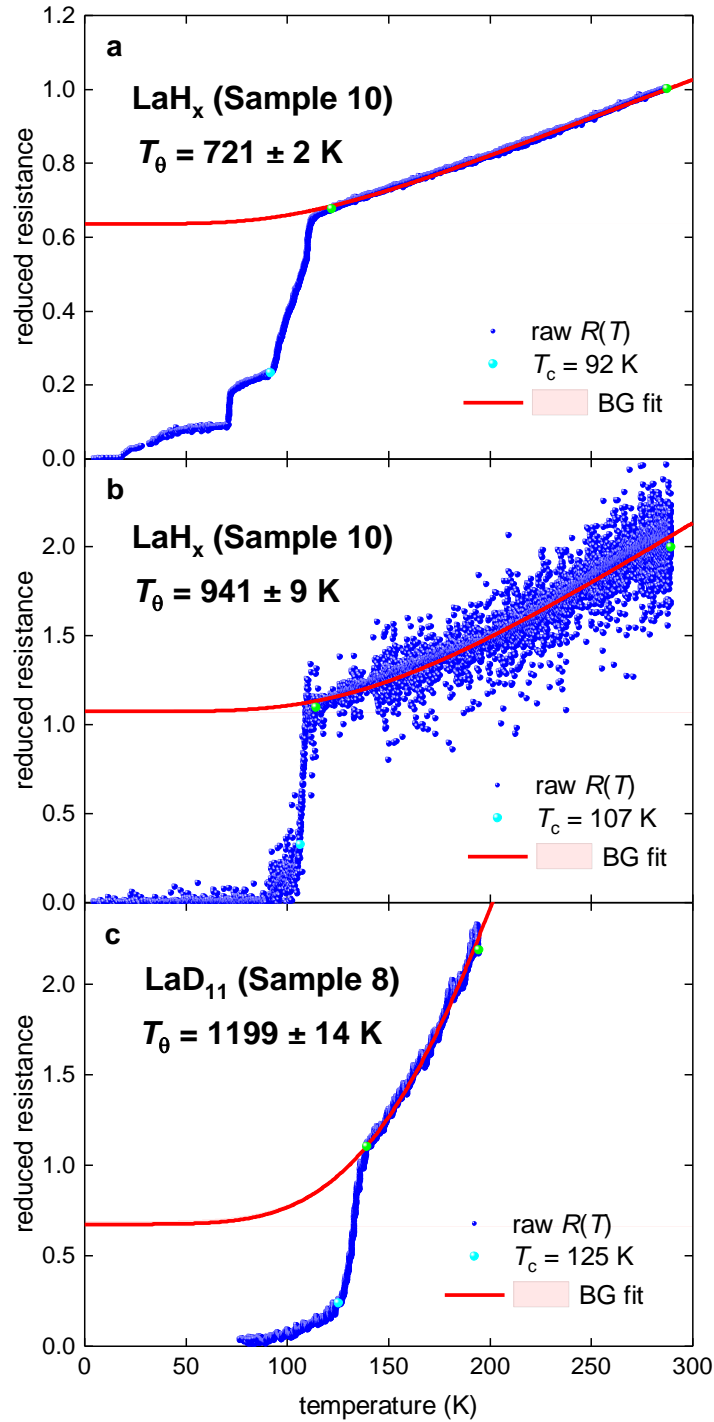

**Figure 7.** Temperature dependent reduced resistance data and fit to BG model for highly-compressed $LaH_x$ (Sample 10 [16], panels **a,b**) and $LaD_{11}$ (Sample 8 [16], panel **c**). $LaH_x$ samples in panels **a** and **b** are at different laser annealing stages. Raw data is from Ref. [16]. 95% confidence bars are shown. Green balls indicate the bounds for which resistance data is used for the fit. Cyan ball shows $T_c$ defined by criteria described in main text. (a) Fit quality $R = 0.9990$; (b) $R = 0.944$; (c) $R = 0.992$.

**Table 5.** Deduced $T_\theta$ and calculated $\lambda_{e\text{-ph,aMcM}}$ for highly-compressed LaH$_x$ (Sample 10 [16]) and LaD$_{11}$ (Sample 8 [16]). Assumed $\mu^* = 0.10$ [34]. $T_c$ values are defined by criteria described in the text.

| Compound (Sample ID; pressure) | $T_\theta$ (K) | $T_c$ (K) | $\lambda_{e-ph,BCS}$ | $\lambda_{e-ph,aMcM}$ | $\lambda_{e-ph}$ (first-principles calculations) [34] |
|---|---|---|---|---|---|
| LaH$_x$ (panel **a**, Fig. 6) (Sample 10; $P = 178$ GPa) | 721 ± 2 | 92 | 0.586 ± 0.001 | 1.82 ± 0.01 | 2.06-2.76 |
| LaH$_x$ (panel **b**, Fig. 6) (Sample 10; $P = 178$ GPa) | 941 ± 9 | 107 | 0.560 ± 0.002 | 1.62 ± 0.01 | 2.06-2.76 |
| LaD$_{11}$ (panel **c**, Fig. 6) (Sample 8; $P = 142$ GPa) | 1199 ± 14 | 125 | 0.51 ± 0.04 | 1.49 ± 0.02 | 3.14 |

### 3.2.5. Other LaH$_x$ and LaD$_y$ samples

Drozdov *et al* [16] in their Extended Data Fig. 5 reported $R(T)/R_{\text{norm}}(T)$ curve for laser annealed LaD$_y$ (Sample 13) specimen compressed at $P = 152$ GPa, which has broad superconducting transition. Transition temperature can be estimated to be about $T_c = 125$ K if the inflection point criterion (Extended Data Fig. 5 [16]) will be applied. However, the fit of $R(T)/R_{\text{norm}}(T)$ to BG equation is not converged, and we were not able to report $T_\theta$ and $\lambda_{e\text{-ph}}$ values for this sample herein.

For similar problems, $T_\theta$ and $\lambda_{e\text{-ph}}$ cannot be deduced for two isotopic counterpart samples with highest reported transition temperatures, i.e. LaH$_{10}$ (Sample 1, $T_c \sim 250$ K) and LaD$_{10}$ (Sample 17, $T_c \sim 180$ K) [16].

### 3.2.6. Overall discussion

For LaH$_{10}$ (Sample 3) which exhibits one of the highest reported transition temperature of $T_c = 240$ K and for which normal part of $R(T)/R_{\text{norm}}(T)$ curve was reported for reasonably wide range of temperatures, we have found remarkably good agreement between $\lambda_{e\text{-ph}}$ values computed by the first principles calculations [34] and deduced from analysis of experimental $R(T)/R_{\text{norm}}(T)$ curve reported herein.

Nevertheless, current approach to compute $\lambda_{\text{e-ph}}$ and $T_c$ in phonon mediated superconductors is based on Allen-Dynes approach which involves the concept of the logarithm of the phonon frequency, $\omega_{\text{ln}}$, which has multiplicative term of $ln(\omega)$. Herein we note, that that latter cannot be taken because the frequency is not unitless value, and logarithm function cannot be taken from a value which has physical unit of Hz.

Taking in account that there is a large disagreement between computed and deduced $\lambda_{\text{e-ph}}$ values for H$_3$S and D$_3$S [22], we can make reaffirm our previous statement [22] that it is more likely that NRT superconductivity is originated from more than one pairing mechanism.

## 4. Conclusion

In this paper we deduce the Debye temperature, $T_\theta$, for all available to date experimental temperature dependent resistance data for LaH-LaD superconductor system [16], for which superconducting transition temperatures varied from 70 K to 240 K. In overall (except of one sample with highest $T_c$), we found a large disagreement between the electron-phonon coupling strength parameter, $\lambda_{\text{e-ph}}$, which we deduced from temperature dependent resistance data and $\lambda_{\text{e-ph}}$ computed by the first principles calculations studies reported by Errea *et al* [34].

**Acknowledgement**

Author thanks financial support provided by the state assignment of Minobrnauki of Russia (theme "Pressure" No. AAAA-A18-118020190104-3) and by Act 211 Government of the Russian Federation, contract No. 02.A03.21.0006.